\documentclass[aps,prb,twocolumn,showpacs,preprintnumbers]{revtex4}
\usepackage{graphicx}
\usepackage{amsmath}
\usepackage{multirow}
\usepackage{comment}
\usepackage[caption=false]{subfig}
\usepackage{tabularx}
\usepackage{adjustbox}

\begin{document}
\title{Survey of the class of isovalent antiperovskite alkaline earth-pnictide compounds}
\author{Wen Fong Goh and Warren E. Pickett }
\affiliation{Department of Physics, University of California, Davis CA 95616, USA}

\begin{abstract}
The few reported members of the antiperovskite structure class $Ae_3Pn_APn_B$ of alkaline earth 
($Ae$ = Ca,Sr,Ba) pnictides ($Pn$ = N,P,As,Sb,Bi) compounds are all based on
the B-site anion $Pn_B$=N. All fit can be categorized as narrow gap semiconductors,
making them of interest for several reasons. Because chemical reasoning
suggests that more members of this class may be stable,
we provide here a density functional theory (DFT)  based survey of this 
entire class of $3\times5\times5$
compounds. We determine first the relative energetic stability of the distribution of
pairs of $Pn$ ions in the A and B sites of the structure, finding that the $B$ site always
favors the small pnictogen anion. The trends of the calculated energy gaps
with $Ae$ cation and $Pn$ anions are determined, and we study 
effects of spin-orbit coupling as well as two types of gap corrections to the conventional 
DFT electronic spectrum. 
Because there have been  suggestions that this class harbors topological
insulating phases, we have given this possibility attention and found that energy
gap corrections indicate the cubic structures will provide at most a few topological
insulators. Structural instability is addressed by calculating phonon dispersion curves
for a few compounds, with one outcome being that distorted structures
should be investigated further for thermoelectric and topological character. 
Examples of the interplay between spin-orbit coupling and strain on the 
topological nature are provided. 
A case study of $\mathrm{Ca_3BiP}$ including the effect of strain illustrates how
a topological semimetal can be transformed into topological 
insulator and Dirac semimetal.
\end{abstract}

\date{\today}
\maketitle

\section{Introduction}
Oxide-based perovskites are the most studied class of ternary compounds 
for several reasons. First, oxygen is abundant and a great deal is known
about oxide chemistry. Second, the basic cubic structure ABO$_3$ is 
simple and is governed
to a great extent by ionic bonding. Third, given the near universal
valence state $O^{-2}$, the $A$ and $B$ cation valences must total six,
leaving a large number of combinations 1:5, 2:4, 3:3, 4:2, 5:1, 
considering the number of atoms that can assume these formal valences. 
Most importantly perhaps is the vast range of properties displayed by
oxide perovskites, including high temperature superconductivity, 
much unusual magnetism, charge, spin, and orbital ordered phases,
large linear responses (viz. Born effective charges that can differ
by a factor of two or more from the formal charge), multiferroic
states, among others. 

Halide-based perovskites, viz. KMnF$_3$ with a monovalent anion, 
are a closely related class of materials. They are
much less prevalent because only 1:2 and 2:1 cation charge pairs are
possible, and the higher eletronegativity makes synthesis less
straightforward. Less common still are nitrogen based perovskites,
because the $N^{-3}$ ion requires cations with formal charge pairs
4:5 or 3:6; higher formal valences are rare. N-based perovskites
have been the topic of a high-throughout computational study to determine
stable examples.\cite{Perez2015}

Within a given crystal structure it is uncommon, but possible, to 
interchange the anion and cation sites. The {\it antiperovskite}
class discussed here will be denoted $Ae_3$AB, where $A$ is the 
(alkaline earth) cation on the oxygen
site and A and B are anions that provide charge balance. Anion
formal valences are practically limited to 1, 2, and 3, constraining
considerably the cation valences. In this paper we will focus on
divalent alkaline earth cations, for which the anion formal charges
must sum to six, so it can be expected that nearly all of such
antiperovskite compounds will have two trivalent anions.   Here we
consider the class in which both A and B anions are from the pnictogen
column of the period table. Given the formal charge balance between
divalent cations and trivalent anions, this class can be anticipated
to harbor insulators and perhaps semimetals. Sun and coauthors have
proposed that topological insulating phases may arise in the
$Ae_3$BiN group with application of uniaxial strain,\cite{Sun2010}
providing additional current interest.

Examples of alkaline earth-pnictide antiperovskite compounds 
have been reported.
Some crystallize in cubic structure,\cite{chern1992a} while others show structural 
distortions\cite{chern1992b} or even form distinct structures, {\it e.g.} 
the two compounds Ba$_3$(Sb,Bi)N assume the hexagonal 
antiperovskite structure\cite{niewa2013}, suggested to be 
due to the presence of large radii alkaline earth ions.
The electronic state of this class of materials is insulating,
ranging from relatively wide gap 
insulating to narrow gap semiconducting. The reported compounds will
be discussed in Sec. \ref{reported_apv} in relation to our results.
Notably, compounds in this class with pnictides other than N inside the
octahedron have not been reported.

\begin{figure}
\subfloat[Tetragonal crystal structure of strained $\mathrm{Ca_3BiP}$.]
  {\includegraphics[width=0.1\linewidth]{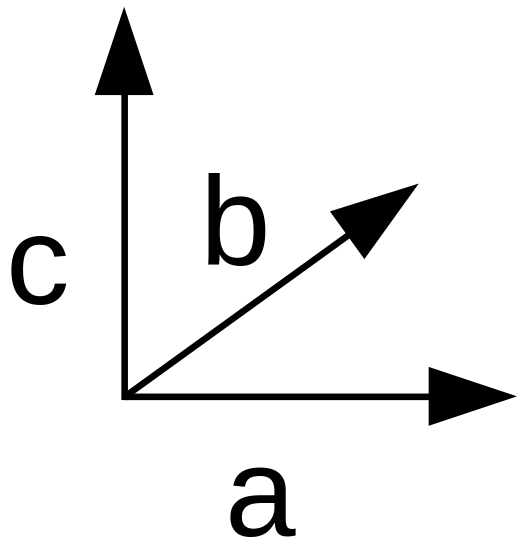}
\includegraphics[width=0.4\linewidth]{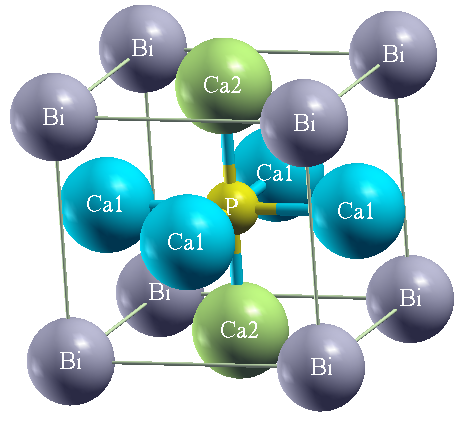}\label{crystal}}
\
\subfloat[Band structure and density of states of cubic $\mathrm{Ca_3BiP}$.]
 {\includegraphics[width=\linewidth]{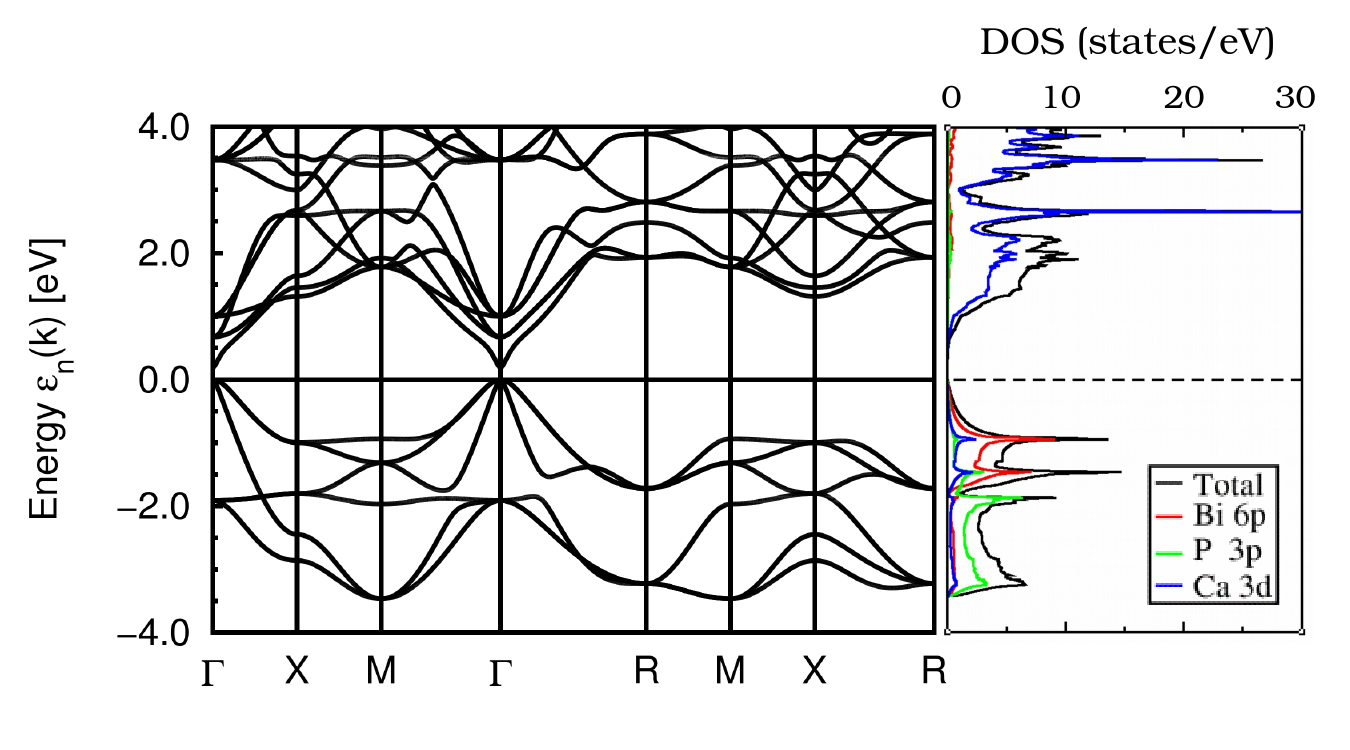}\label{band}}
\caption{
(a) The antiperovskite structure of $\mathrm{Ca_3BiP}$ with tetragonal compression
along (001), the space group becomes $P4/mmm$. Ca$^{2+}$ ions lie on two nonequivalent sites(blue and green).
(b) Band structure and density of states of cubic $\mathrm{Ca_3BiP}$, without SOC.
P $3p$ bands provide the lowest valence states, while Bi $6p$ bands comprise
the highest valence states. Ca $4s$ and $3d$ orbitals dominate the lower conduction states.
The gap of 0.2 eV occurs at $\Gamma$, throughout the rest of the zone the gap is large.}
\end{figure}

Density functional theory (DFT) based study of insulating Ca$_3$BiN
and metallic Ca$_3$PbN provided the underlying electronic structure
(as shown in the current paper) of this class of materials.\cite{papa1992}
DFT based results for the band structure and selected properties
(primarily bandgap, elastic, thermoelectric)
have been reported for several of
these compounds, which are discussed in Sec. \ref{reported_apv} below.
Motivated by the relatively few known examples and the 
broad possibilities for new materials with
important properties, we have carried out a survey on the entire class of $3\times5\times5$  
alkaline earth-pnictide antiperovskite compounds, viz. $Ae_3Pn_APn_B$, 
where $Ae = \mathrm{Ca, Sr, Ba}$ and $Pn_A, Pn_B = \mathrm{N, P, As, Sb, Bi}$, using first 
principles DFT methods.
The conventional ordering of atoms $Pn_A$ and $Pn_B$ will follow 
that of a perovskite with general chemical formula $\mathrm{ABO_3}$, 
where the $\mathrm{A}$ cation is 12-fold cub-octahedral coordinated and $\mathrm{B}$ is 
6-fold coordinated by an octahedron of O anions.
In the case of antiperovskite $Ae_3Pn_APn_B$, $Pn_B$ is inside the 
$Ae_6$ octahedron while $Pn_A$ sits in the more open A site.

The manuscript is organized as follows. The computational methods are summarized
is Sec. \ref{method}.  Section \ref{analysis} contains much of the basic results, and includes a
general analysis of the common features of these antiperovskites, the correction
to DFT bandgaps that are important for small (or negative) gap compounds, and a
synopsis of previously reported or predicted members of this class. The possibility of 
topological character, and the favorable candidates, are described Sec. \ref{topology}.
Section \ref{structure} addresses the question of structural stability with two examples, and
a concise summary is given in Sec. \ref{summary}.

\section{Computational Methods}
\label{method}

To study the electronic structure of these 75 compounds, the generalized gradient approximation 
(GGA) exchange-correlation functional of Perdew-Burke-Ernzerhof-1996\cite{Perdew1996} was
used, as implemented in the full-potential local orbital (FPLO) \cite{Koepernik1999} scheme.
Self-consistency was obtained on a dense $k$-mesh of $20 \times 20 \times 20$ to ensure 
good convergence of the density and thereby the eigenvalues.
The fully relativistic Dirac four component equations implemented in FPLO were performed
to include spin-orbit coupling and other relativistic corrections. The cubic lattice
constants were obtained by minimizing the energy with respect to volume.

In small gap compounds, and especially for potential topological materials where
gaps are typically quite small, the underestimate of the band gap by the GGA functional
must be dealt with. 
We have calculated band gap corrections of the $Ae_3$Bi$Pn_B$ subclass using the 
modified Becke-Johnson (mBJ) exchange-correlation potential. This scheme provides a
self-energy-like correction to eigenvalues that has been found to be realistic for
several semiconductors and insulators.\cite{Becke2006,Tran2007,Tran2009} A 
formally well justified
means of obtaining self-energy corrections is provided by 
GW approximation.\cite{FHI-gap,Jiang2016} 
We have used the FHI-gap implementation\cite{FHI-gap} 
in the WIEN2K code\cite{Blaha2001} on four compounds,
and make comparison with the mBJ results in Sec. \ref{gap_correction}. 
In this $GW_{\circ}$
implementation the eigenvalue in $G_{\circ}$ is iterated to self-consistency,
hence our use of the terminology $GW_{\circ}$ throughout.
The linearized augmented plane wave basis was specified by $R_m G_{max}$=7, 
where $R_m$ in the minimum radius of the atomic spheres and $G_{max}$ is
the largest reciprocal lattice vector. For the $GW_{\circ}$ version of
$GW$ implemented in FHI-gap, the mixed basis quality was determined by
$\ell_{max}$ = 3, $Q$ = 0.75. Unoccupied states up to 1000 eV were summed over,
the number of discrete imaginary frequency points was 16,
and the k-point mesh was $3\times 3\times 3$. Increasing the latter to 
$4\times 4\times 4$ resulted in self-energy changes of the order of 0.01 eV.

\section{Analysis of Cubic Structures and Spin-Orbit Coupling}
\label{analysis}

\begin{figure}
\centering
\includegraphics[width=0.5\textwidth]{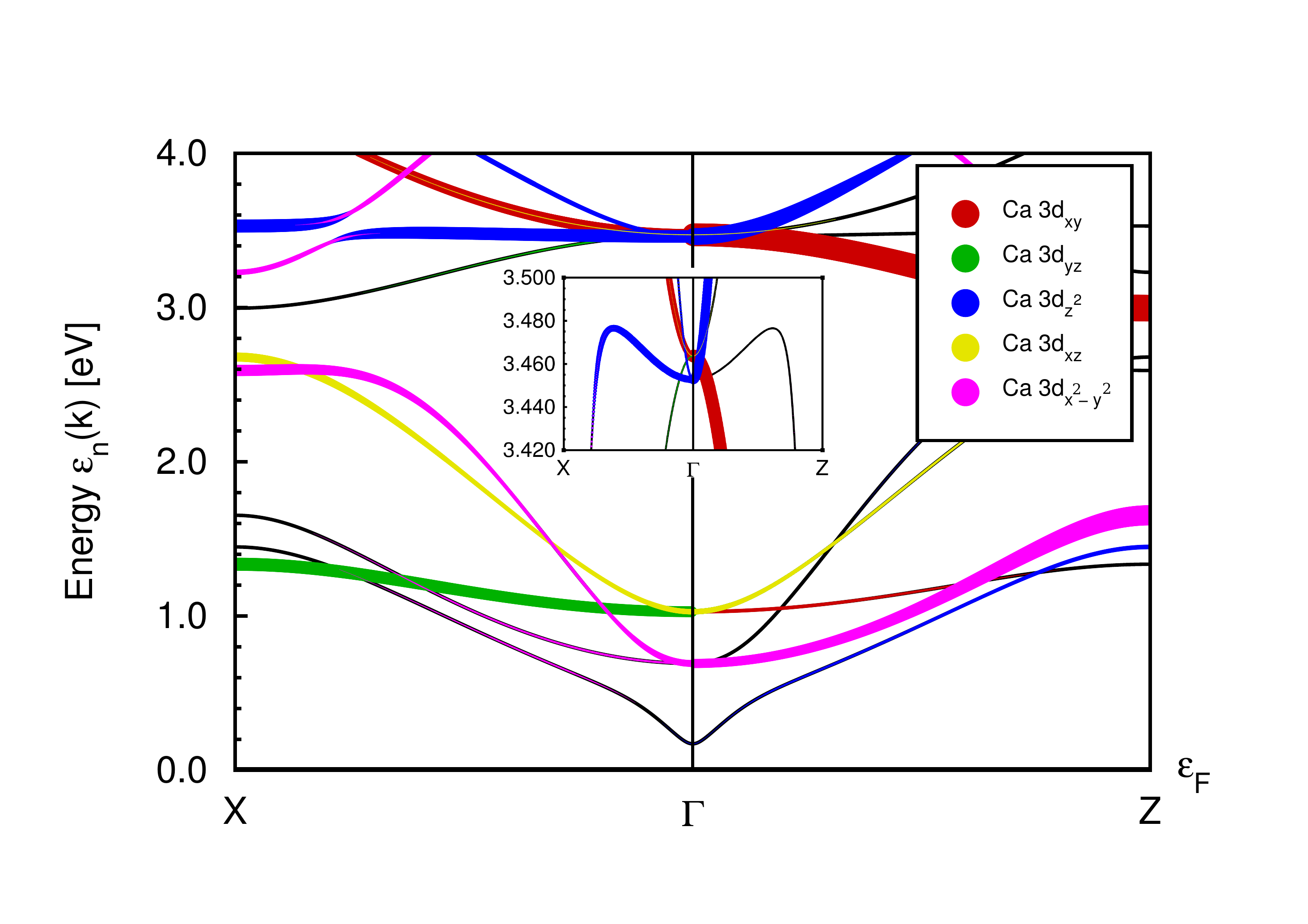}%
\caption{Color plot differentiating orbital character, showing the band ordering 
of $3d$ orbital energies of the apical Ca ions located at
$(\frac{1}{2},\frac{1}{2},0)a$ in the unit cell.
The insert provides an enlargement of the complex near $\Gamma$ around 3.5 eV,
showing that the  $d_{xy}$ and $d_{z^2}$ levels are nearly degenerate at $\Gamma$.}
\label{ca}
\end{figure}

\subsection{Cell volume and energetic stability}

\begin{table*}[]
\caption{Data on the complete class of $Ae_3Pn_APn_B$ antiperovskites.
$a(\mathrm{\AA})$ gives the optimized lattice constants in \AA. 
$\Delta E = E_{Ae_3Pn_APn_B} - 
   E_{Ae_3Pn_BPn_A}$ provides the energy difference in eV. Since this
energy difference is antisymmetric across the diagonal, the energy differences
are only shown in the lower left triangle for the stable phase (the smaller $Pn$
ion is always $Pn_B$).
$\varepsilon_g$ and $\varepsilon_g^{SOC}$ give the band gap 
(without and with SOC) in eV; of either a small positive 
or negative (negative indirect band gap) value, 
with an exception of a few with bands crossing at the Fermi level 
or gapless states (represented by SM for semimetallic or M for metallic state).
Only the $\nu_0$ $Z_2$ index is provided, because all $\nu_{1,2,3}$ are zero. 
Values in bold indicate a topological semimetal state.}
\label{table1}
\resizebox{2\columnwidth}{!}{
\begin{tabular}{|c|c|c|c|c|c|c|c|c|c|c|c|c|c|c|c|c|c|c|c|c|c|c|c|c|c|c}
\hline
                        & \multicolumn{6}{c|}{N$_B$}                                                                      & \multicolumn{5}{c|}{P$_B$}                                                                      & \multicolumn{5}{c|}{As$_B$}                                                                     & \multicolumn{5}{c|}{Sb$_B$}                                                                     & \multicolumn{5}{c|}{Bi$_B$}                                                                                     \\ \hline
\multirow{4}{*}{N$_A$}  &    & a(\AA) & $\Delta E$(eV) & $\varepsilon_g$(eV) & $\varepsilon_g^{SOC}$(eV) & $Z_2$ & a(\AA) & $\Delta E$(eV) & $\varepsilon_g$(eV) & $\varepsilon_g^{SOC}$(eV) & $Z_2$      & a(\AA) & $\Delta E$(eV) & $\varepsilon_g$(eV) & $\varepsilon_g^{SOC}$(eV) & $Z_2$      & a(\AA) & $\Delta E$(eV) & $\varepsilon_g$(eV) & $\varepsilon_g^{SOC}$(eV) & $Z_2$      & a(\AA) & $\Delta E$(eV) & $\varepsilon_g$(eV) & $\varepsilon_g^{SOC}$(eV) & \multicolumn{1}{c|}{$Z_2$} \\ \cline{2-27} 
                        & Ca & 4.61   & -                       & 0.                  & 0.                        & 1     & 5.29   & -                       & SM                  & M                         & 1          & 5.42   & -                       & 0.                  & 0.                        & 1          & 5.76   & -                       & -0.01               & -0.02                     & 1          & 5.85   & -                       & -0.01               & -0.04                     & \multicolumn{1}{c|}{1}     \\ \cline{2-27} 
                        & Sr & 4.99   & -                       & 0.                  & 0.                        & 1     & 5.67   & -                       & 0.                  & 0.                        & 1          & 5.80   & -                       & 0.                  & -0.01                     & 1          & 6.13   & -                       & -0.02               & -0.03                     & 1          & 6.23   & -                       & -0.02               & -0.05                     & \multicolumn{1}{c|}{1}     \\ \cline{2-27} 
                        & Ba & 5.33   & -                       & SM                  & SM                        & 1     & 6.03   & -                       & -0.04               & -0.05                     & 1          & 6.18   & -                       & -0.05               & -0.06                     & 1          & 6.52   & -                       & -0.07               & -0.08                     & 1          & 6.61   & -                       & -0.05               & -0.12                     & \multicolumn{1}{c|}{1}     \\ \hline
\multirow{3}{*}{P$_A$}  & Ca & 4.73   & -2.96                   & 0.85                & 0.85                      & 0     & 5.31   & -                       & 0.33                & 0.31                      & 0          & 5.42   & -                       & 0.10                & 0.05                      & 0          & 5.74   & -                       & 0.                  & 0.                        & 1          & 5.82   & -                       & 0.                  & 0.                        & \multicolumn{1}{c|}{1}     \\ \cline{2-27} 
                        & Sr & 5.09   & -2.55                   & 0.45                & 0.45                      & 0     & 5.67   & -                       & 0.                  & 0.                        & 1          & 5.79   & -                       & 0.                  & 0.                        & 1          & 6.09   & -                       & 0.                  & 0.                        & 1          & 6.17   & -                       & 0.                  & 0.                        & \multicolumn{1}{c|}{1}     \\ \cline{2-27} 
                        & Ba & 5.43   & -2.28                   & SM                  & SM                        & 0     & 6.02   & -                       & 0.03                & 0.01                      & 1          & 6.14   & -                       & 0.                  & 0.                        & 1          & 6.44   & -                       & 0.                  & 0.                        & 1          & 6.53   & -                       & 0.                  & -0.10                     & \multicolumn{1}{c|}{1}     \\ \hline
\multirow{3}{*}{As$_A$} & Ca & 4.78   & -3.38                   & 0.77                & 0.72                      & 0     & 5.34   & -0.55                   & 0.15                & 0.09                      & 0          & 5.45   & -                       & 0.                  & 0.                        & 1          & 5.75   & -                       & 0.                  & 0.                        & 1          & 5.84   & -                       & 0.                  & 0.                        & \multicolumn{1}{c|}{1}     \\ \cline{2-27} 
                        & Sr & 5.13   & -2.93                   & 0.30                & 0.28                      & 0     & 5.70   & -0.50                   & 0.                  & 0.                        & \textbf{1} & 5.81   & -                       & 0.                  & 0.                        & 1          & 6.11   & -                       & 0.                  & 0.                        & 1          & 6.19   & -                       & 0.                  & -0.01                     & \multicolumn{1}{c|}{1}     \\ \cline{2-27} 
                        & Ba & 5.47   & -2.64                   & SM                  & SM                        & 0     & 6.04   & -0.46                   & 0.                  & 0.                        & \textbf{1} & 6.16   & -                       & 0.                  & 0.                        & 1          & 6.46   & -                       & 0.                  & -0.02                     & 1          & 6.54   & -                       & 0.                  & -0.14                     & \multicolumn{1}{c|}{1}     \\ \hline
\multirow{3}{*}{Sb$_A$} & Ca & 4.88   & -4.26                   & 0.46                & 0.35                      & 0     & 5.39   & -1.76                   & 0.65                & 0.50                      & 0          & 5.49   & -1.2                    & 0.40                & 0.22                      & 0          & 5.78   & -                       & 0.07                & 0.                        & 1          & 5.86   & -                       & 0.                  & 0.                        & \multicolumn{1}{c|}{1}     \\ \cline{2-27} 
                        & Sr & 5.22   & -3.77                   & 0.20                & 0.15                      & 0     & 5.74   & -1.58                   & 0.30                & 0.16                      & 0          & 5.84   & -1.08                   & 0.10                & 0.                        & \textbf{1} & 6.13   & -                       & 0.                  & 0.                        & 1          & 6.21   & -                       & 0.                  & 0.                        & \multicolumn{1}{c|}{1}     \\ \cline{2-27} 
                        & Ba & 5.55   & -3.40                   & SM                  & SM                        & 0     & 6.08   & -1.45                   & 0.28                & 0.18                      & 0          & 6.19   & -0.99                   & 0.13                & 0                         & \textbf{1} & 6.48   & -                       & 0.                  & 0.                        & 1          & 6.55   & -                       & 0.                  & -0.12                     & \multicolumn{1}{c|}{1}     \\ \hline
\multirow{3}{*}{Bi$_A$} & Ca & 4.92   & -4.35                   & 0.48                & 0.07                      & 0     & 5.42   & -1.99                   & 0.16                & 0.                        & \textbf{1} & 5.53   & -1.44                   & 0.                  & -0.03                     & \textbf{1} & 5.81   & -0.29                   & 0.                  & -0.10                     & \textbf{1} & 5.89   & -                       & 0.                  & -0.12                     & \multicolumn{1}{c|}{0}     \\ \cline{2-27} 
                        & Sr & 5.26   & -3.86                   & 0.25                & 0.01                      & 0     & 5.77   & -1.75                   & 0.                  & -0.02                     & \textbf{1} & 5.88   & -1.26                   & 0.                  & -0.05                     & \textbf{1} & 6.15   & -0.20                   & 0.                  & -0.10                     & \textbf{1} & 6.23   & -                       & 0.                  & -0.13                     & \multicolumn{1}{c|}{0}     \\ \cline{2-27} 
                        & Ba & 5.59   & -3.47                   & SM                  & SM                        & 0     & 6.11   & -1.61                   & 0.                  & -0.05                     & \textbf{1} & 6.22   & -1.17                   & 0.                  & -0.10                     & \textbf{1} & 6.50   & -0.21                   & 0.                  & -0.18                     & \textbf{1} & 6.58   & -                       & 0.                  & -0.17                     & \multicolumn{1}{c|}{0}     \\ \hline
\end{tabular}
}
\end{table*}

The cubic antiperovskite $Ae_3Pn_APn_B$ structure \cite{niewa2013} 
(space group $Pm\bar{3}m$) has an $Ae_6Pn_B$ octahedron, with $Pn_A$ 
surrounded symmetrically by eight octahedra, with the primitive cell
pictured in Fig. \ref{crystal}.
For the cubic (undistorted) structure,
the equilibrium lattice constant has been obtained for all 
$3\times 5\times 5$ combinations. These results and the following data
are presented in Table \ref{table1}. First is a comparison of the energy difference
upon  interchange $Pn_A \leftrightarrow Pn_B$ of the two $Pn$ ions. 
It is always the case that it is
energetically favorable to have the smaller $Pn$ ion in the $Pn_B$ octahedron
position. Energy differences, which range from 0.2-4.3 eV,
are largest when N is one of the ions, and the
magnitude of the difference increases with difference in atomic number
of the anions.
These trends can be understood from Coulomb energetics:
if the smaller anion is surrounded by the cation octahedron, the 
attractive Coulomb energy will be larger. 
An additional factor is that the A site naturally accommodates a bigger
atom than will the B site, giving better volume filling by atoms.
The other data in Table \ref{table1} will be discussed below.

\subsection{Electronic band structure}

For general orientation, the electronic structure and DOS of a representative
compound, Ca$_3$BiP containing the largest and the next to smallest pnictide ions, is shown in
Fig. \ref{band}.
A generic feature in this class of perovskites is that the smaller $Pn_B$  $p$ bands
lie lower than the $Pn_A$ $p$ bands, making the larger ion (which conveniently also has
the largest SOC) the one of interest in
determining the band gap (or not) and subsequently the topological character.
Since both A and B sites have cubic symmetry, both $Pn$ $p$ band complexes have
threefold degeneracy at the $\Gamma$ point when SOC is neglected, and SOC separates
the eigenvalues into $p_{1/2}$ and $p_{3/2}$ states, with the
latter for $Pn_B$ forming the top of the valence bands.

In the conduction bands three sets of five $Ae$ $d$ bands dominate. Unexpectedly, a
``free electron like'' band that has no single dominant orbital character, discussed below,
lies near the bottom of the conduction bands at $\Gamma$ and often is the band at
the conduction band minimum (CBM).
Since the $Ae$ site has tetragonal symmetry, the $d$
shell is split by the crystal field into a $d_{xz},d_{yz}$ doublet and three singlets
$d_{xy}, d_{x^2-y^2}, d_{z^2}$, relative to the local ($z$) axis. For some purposes it
would be important to consider how these thirty orbitals on six $Ae$ ions form
molecular orbitals on the $Pn_BAe_6$ octahedron. Actually performing and applying
such a transformation is complicated by the fact that each $Ae$ ions' orbitals are
shared with two octahedra.

One fundamental interaction at and near $\Gamma$ is that of the $Pn_A$ $p$ states
with the $d$ orbitals of the twelve $Ae$ ions coordinating it, and with the
diffuse ``$s$'' band.
To assist understanding of the ordering of the $Ae$ $d$ characters at the $\Gamma$ point,
the orbital characters in the
$d$ band region of the apical Ca ion (whose local axis is along the $\hat z$ direction)
are plotted in Fig.~\ref{ca}.
Ca $d_{xy}$ and $d_{z^2}$ have orbital density oriented towards four nearest Bi$^{-3}$ ions
and two nearest P$^{-3}$ ions respectively, and their orbital energies are
close and are higher than for the other $d$ orbitals.
The $d_{x^2-y^2}$ orbital lies lowest in energy, corresponding to an orbital density
directed between negative ions, being lower than the highest $d$ orbital by 2.8 eV.
This separation changes to 3.6 eV and 3.3 eV by replacing Ca with Sr and Ba respectively.
The degenerate $d_{xz}$, $d_{yz}$ orbitals lie about 0.4 eV above the $d_{x^2-y^2}$ orbital.
The $Pn_A$ $p$ bands approach and mix with the $Ae$ $d$ bands around the $\Gamma$ point,
with a substantial band gap throughout the rest of the zone.

{\it Trends in band gaps.}
We here review the behavior of the gaps of the more stable members, which lie in the
lower left triangle of Table I. We consider the bands without SOC, which are simpler
from which to derive chemical trends. For N$_B$ (N on the B site), the Ca compounds have gaps
that decrease from 0.85 eV to 0.48 eV for the progression from P$_A$ to Bi$_A$. The
corresponding Sr compounds have smaller gaps, 0.45 eV to 0.25 eV, and for the Ba compounds
there is band overlap (negative overlap). Moving to P$_B$ and on to Bi$_B$, there is no monotonic
progression of gaps, but a clear trend to smaller or zero gaps at the B site
atom becomes heavier. Evidently SOC, which splits the valence band maximum (VBM), 
will decrease the gap size and in some cases
lead to gap inversion (reported in the table as zero gap or negative indirect gap).

{\it Topological character.} One of the motivations of this survey of 75 compounds
was to begin to assess the likelihood of finding topological insulating, or also
topological semimetal, phases in these antiperovskites. We return to this topic
in Sec. \ref{topology}.
However, in Table I we have provided the $Z_2$ topological
index for the compounds and structures we have studied, for the GGA+SOC bands.
In the lower left subdiagonal of Table I that contains the more stable ordering of
the $Pn_A, Pn_B$ pairs, we find no topological bands in the N$_B$ column. However,
moving to the heavier $Pn_B$ columns, many $Z_2=1$ results are obtained. The
profusion of these values appear at first sight to promote optimism. There are
important caveats, however. First, GGA calculations are known to underestimate
bandgaps in insulators. Self-energy corrections to the gap are discussed in Sec. \ref{gap_correction}.
Secondly, stability of these model compounds is an issue. Only a few have
been reported, which we discuss in Sec. \ref{reported_apv}. Some may be thermodynamically
unstable with respect to more stable phases, while other may be dynamically
unstable and distort to a lower symmetry structure with an altered
electronic structure. We return to the question of topological character
in Sec. \ref{topology}.

\subsection{Energy Gap Correction}
\label{gap_correction}

\begin{figure*}[ht]
    \centering
    \subfloat[Ca$_3$BiN]{%
        \includegraphics[width=0.4\linewidth]{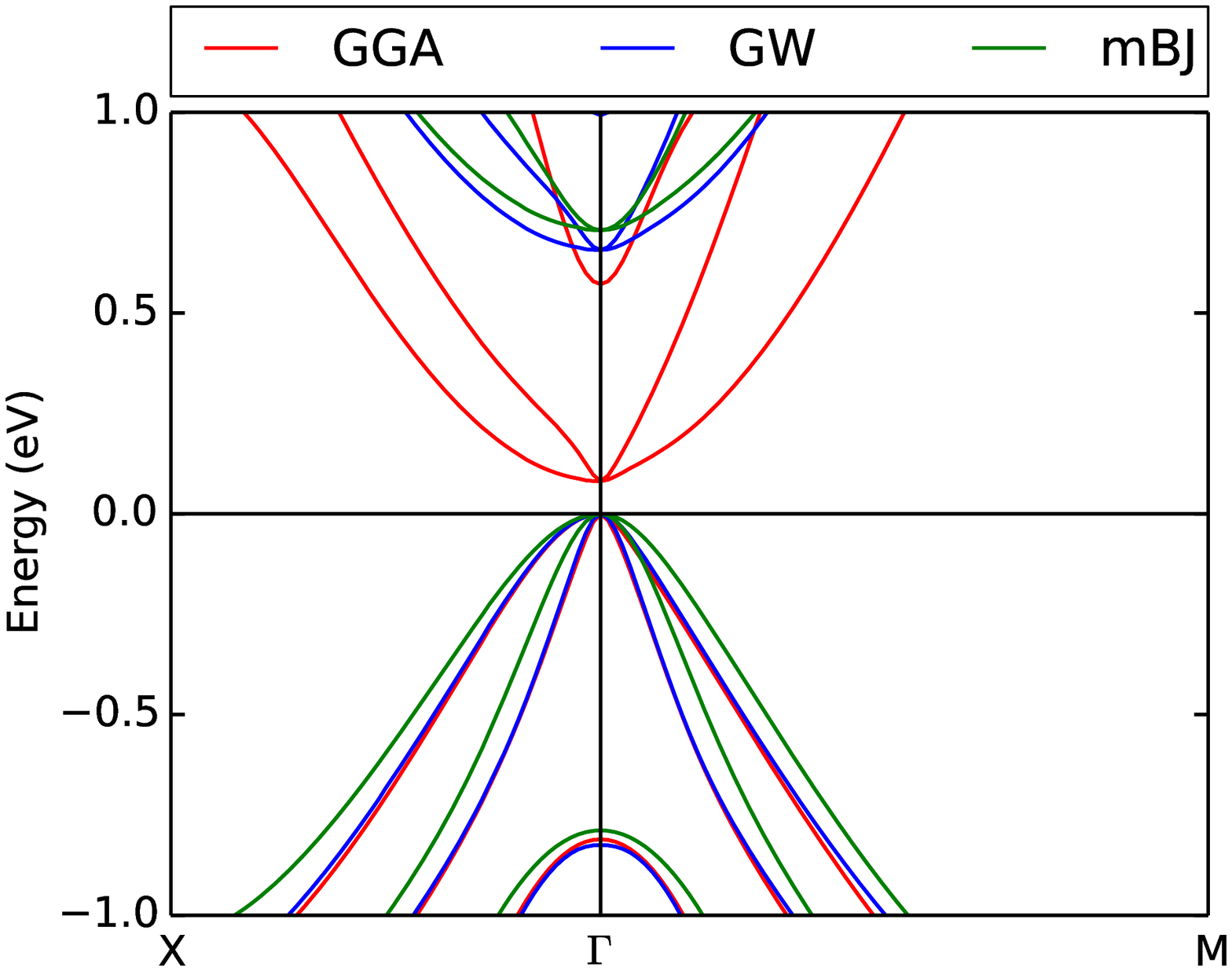}\label{gw-a}%
        }%
    \subfloat[Ca$_3$BiP]{%
        \includegraphics[width=0.4\linewidth]{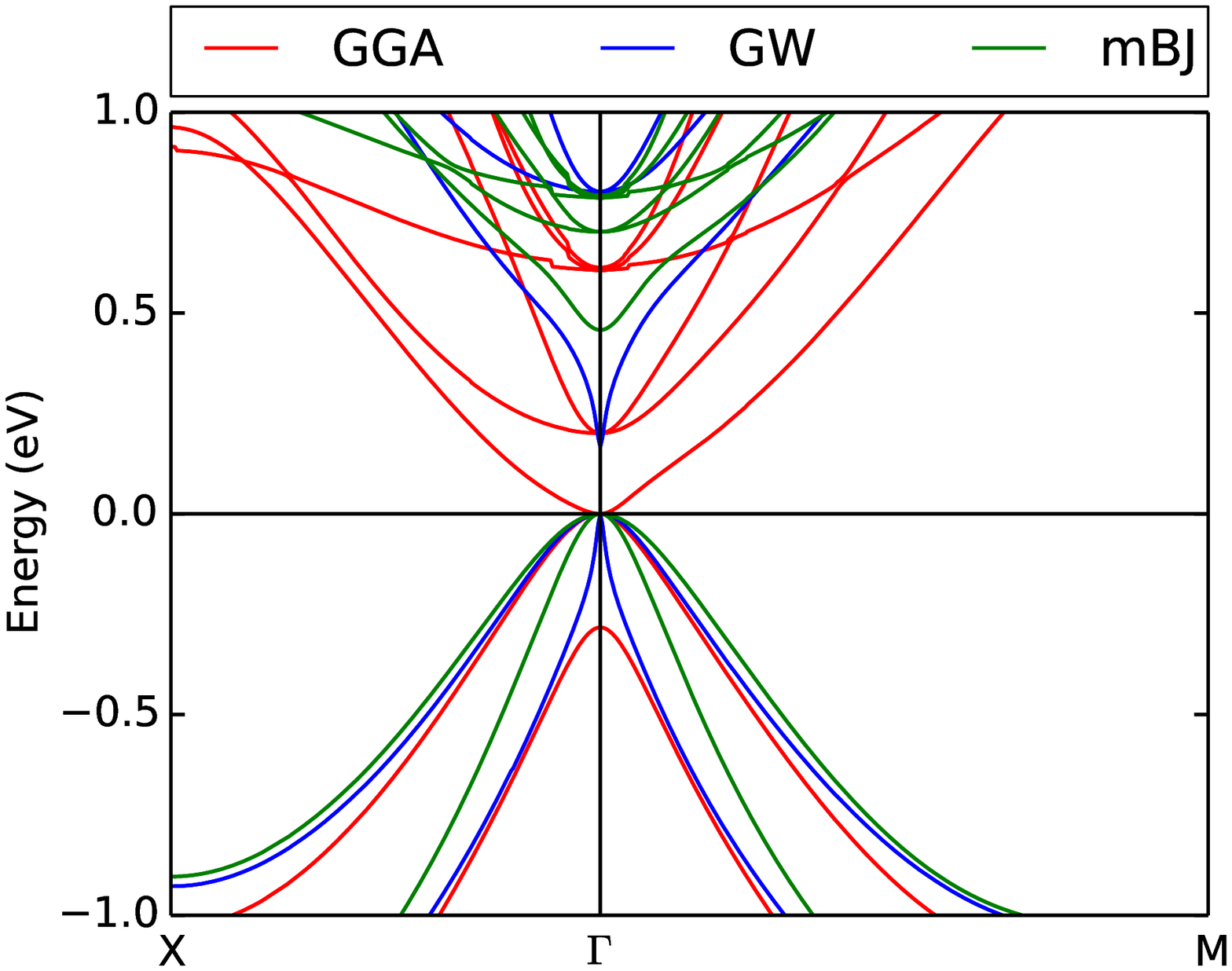}\label{gw-b}%
        }\\%
    \subfloat[Sr$_3$BiN]{%
        \includegraphics[width=0.4\linewidth]{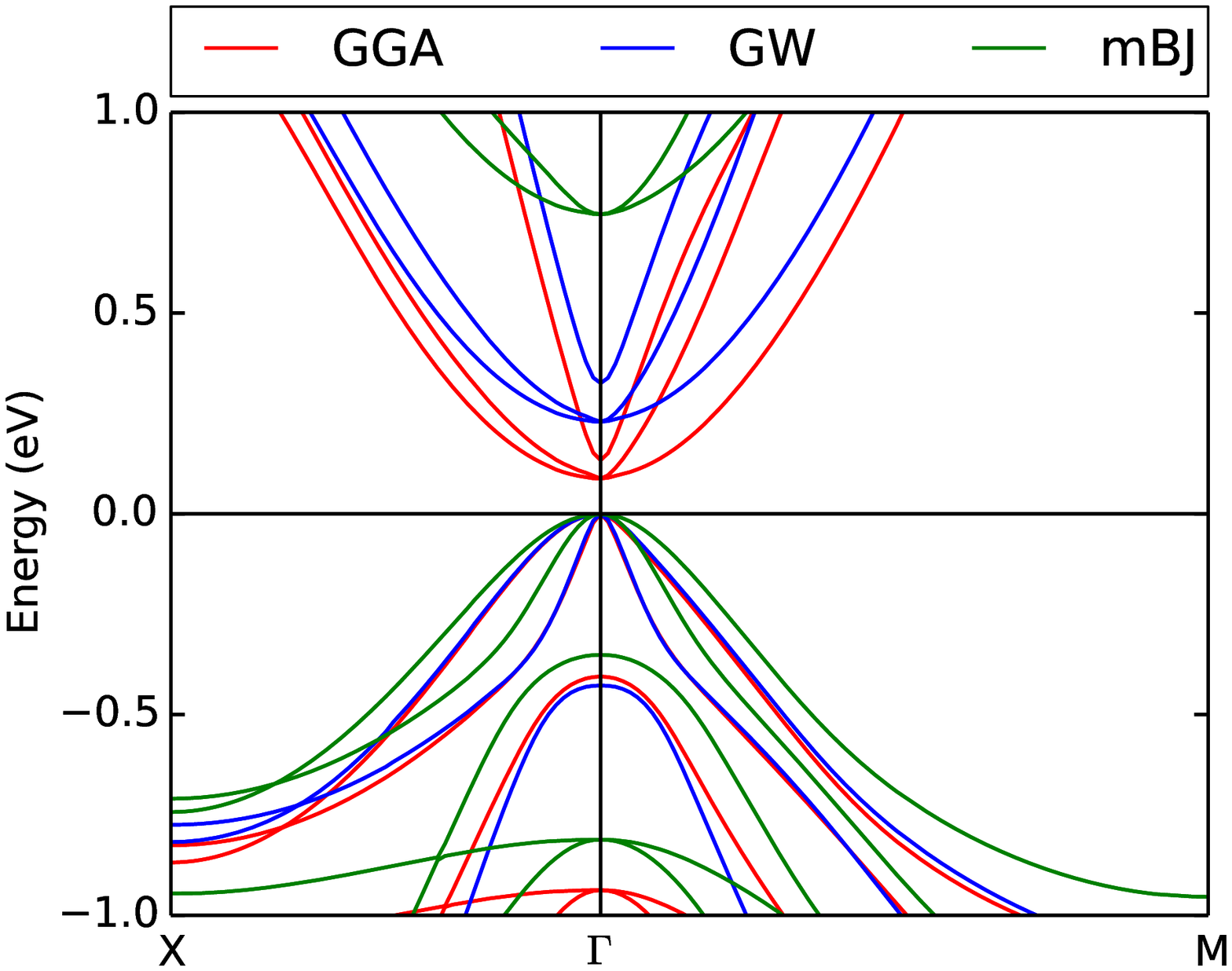}\label{gw-c}%
        }%
    \subfloat[Sr$_3$BiAs]{%
        \includegraphics[width=0.4\linewidth]{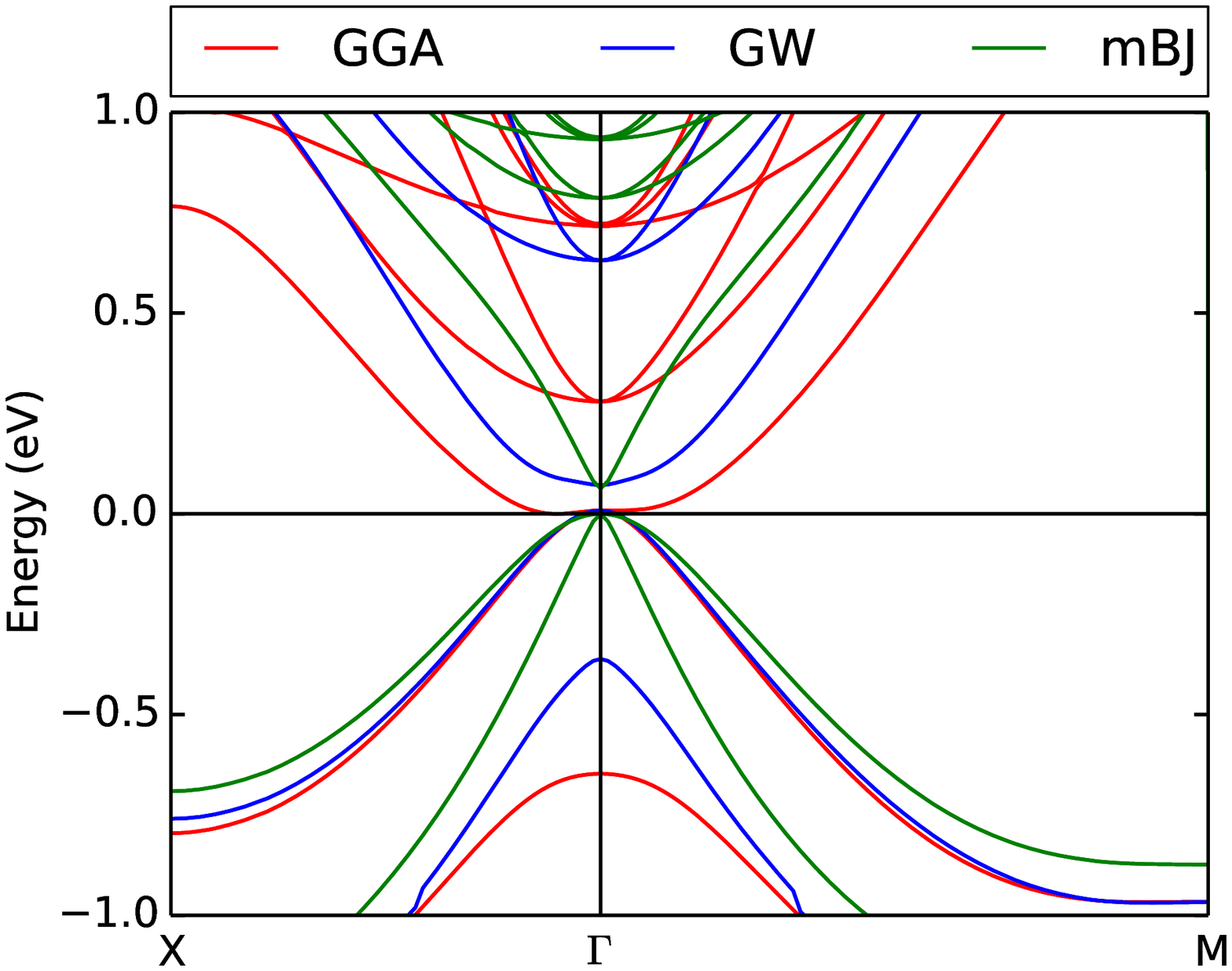}\label{gw-d}%
        }%
    \caption{GGA (red), GW$_{\circ}$ (blue) and mBJ (green) energy bands along 
     X - $\Gamma$ - M directions, for four antiperovskite compounds. 
     All include spin-orbit coupling.}
    \label{gw}
\end{figure*}

We define the inversion energy as the energy of the diffuse ``s'' band
minus the energy of the VBM (which is that of the
A site $p$-band and has negative parity). 
(In the case the B site atom is N, the VBM is a strong
admixture of $Pn_A$ and N $p$ character.)
A negative inversion energy indicates the band inversion that is
essential for topological character.
Table \ref{table2} provides the inversion energy and band gap for 
GGA+SOC eigenvalues,
with and without mBJ. Note that the band gap, once SOC is included,
 usually appears away from the
$\Gamma$ point, thus is not equal to the inversion energy.
The effect of including mBJ is a separation of conduction bands from
valence bands (inversion energy) of 1.0, 0.7, 0.7, 0.6 eV for
N, P, As, Sb, respectively, a very systematic trend independent of
the cation Ca, Sr, Ba.  The final (GGA+mBJ+SOC) band gap, if nonzero, is less
regular because it occurs away from the $\Gamma$ point and is more
dependent of details of the dispersion and inversion energy.

Only the $Ae_3$BiSb compounds show negative inversion energy
when the mBJ correction is included.
$\mathrm{Sr_3BiAs}$ has very small positive inversion energy (0.06 eV).
The rest are normal insulators. Many of these inversion energies, positive
or negative, are small enough that they can be manipulated by pressure
or strain, with an example discussed in Sec. \ref{topology}.

The bands of the four compounds we have chosen are shown in Fig.~\ref{gw},
all being based on Bi on the A site to provide the largest effect of SOC.
Ca$_3$BiN in Fig. \ref{gw-a} provides the reference normal insulator situation,
with a two-fold level (A atom $p_{3/2}$) at the VBM and a doublet
and a somewhat higher singlet in the lower conduction bands. The bands for the three
cases, GGA(SOC), GGA(SOC)+mBJ, GGA(SOC)+GW$_{\circ}$,
are aligned at the VBM. The corrections to the valence bands are not very 
important for our current considerations, but the two corrections are
rather small but not necessarily similar. The conduction band corrections
are larger in the region pictured, about 0.5 eV, and similar in the
low energy region that is shown. Comparison between mBJ and GW corrections
have been commented on previously.\cite{Camargo-Martinez2012} 

Fig. \ref{gw-c} shows the Sr analog, Sr$_3$BiN. Its gap is smaller,
but more interesting is that the conduction band singlet has dropped
close to the doublet. Unexpectedly, the GW$_{\circ}$ correction
has dropped to 0.15 eV while mBJ remains near 0.5 eV. The corrections
in the valence bands remain small, and again the two corrections are
not always similar.

Band inversion occurs for Ca$_3$BiP shown in Fig.~\ref{gw-b}, due to
the conduction band singlet dropping not only below the doublet, but
also below the VBM doublet. One crucial result is that the band
structure becomes that of a zero-gap semiconductor.
Away from $\Gamma$ the bands hybridize, 
complicating the process of following the underlying bands. 
However, the inversion becomes clear
when it is noticed that both self-energy corrections {\it raise} the
$\Gamma$ GGA(SOC) eigenvalue at -0.3 eV, consistent with conduction
eigenvalue character, which is also confirmed by its positive parity
eigenvalue. 

This behavior is repeated for Sr$_3$BiAs shown in Fig.~\ref{gw-d},
where the band orderings and corrections are somewhat different than
in Ca$_3$BiP and may be easier to follow. Even in GGA(SOC) the low
energy band structure can be quite delicate, but becomes more so
in the presence of self-energy corrections. This complex behavior
will be explored on more detail elsewhere.

\begin{table*}[]
\caption{Inversion energy and band gap of $Ae_3\mathrm{Bi}Pn$, where $Ae=\mathrm{Ca,Sr,Ba}$ 
   and $Pn=\mathrm{N,P,As,Sb,Bi}$ with SOC and mBJ correction.  The mBJ effect is defined 
   as the difference between with and without mBJ.}
\label{table2}
\resizebox{\textwidth}{!}{
\begin{tabular}{|c|l|l|l|l|l|l|l|l|l|l|l|l|l|}
\hline
\multicolumn{1}{|l|}{$Ae_3\mathrm{Bi}Pn$} & \multicolumn{1}{c|}{\it{Pn}} & \multicolumn{3}{c|}{N}           & \multicolumn{3}{c|}{P}            & \multicolumn{3}{c|}{As}           & \multicolumn{3}{c|}{Sb}           \\ \hline
\it{Ae}                                         &                         & SOC     & SOC + mBJ & mBJ effect & SOC      & SOC + mBJ & mBJ effect & SOC      & SOC + mBJ & mBJ effect & SOC      & SOC + mBJ & mBJ effect \\ \hline
\multirow{2}{*}{Ca}                       & Inversion Energy        & 0.572 & 1.587   & 1.015    & -0.283 & 0.460   & 0.743    & -0.533 & 0.188   & 0.721    & -0.754 & -0.149  & 0.604     \\ \cline{2-14} 
                                          & Band Gap                & 0.081 & 0.706   & 0.625    & 0        & 0.460   & 0.460    & 0        & 0.188   & 0.188    & 0        & 0         & 0          \\ \hline
\multirow{2}{*}{Sr}                       & Inversion Energy        & 0.133 & 1.138   & 1.005    & -0.458 & 0.286   & 0.744     & -0.658 & 0.063   & 0.721    & -0.831 & -0.226  & 0.605    \\ \cline{2-14} 
                                          & Band Gap                & 0.088 & 0.746   & 0.658    & 0        & 0.286   & 0.286    & 0        & 0.063   & 0.063    & 0        & 0         & 0          \\ \hline
\multirow{2}{*}{Ba}                       & Inversion Energy        & 0.003 & 1.010   & 1.007    & -0.264 & 0.464   & 0.727    & -0.427 & 0.283    & 0.709    & -0.633 & -0.034  & 0.599    \\ \cline{2-14} 
                                          & Band Gap                & 0       & 0.483   & 0.483    & 0        & 0.464   & 0.464    & 0        & 0.283    & 0.283     & 0        & 0         & 0          \\ \hline
\end{tabular}
}
\end{table*}

\subsection{Reported antiperovskites}
\label{reported_apv}

Relatively few of this class of antiperovskite compounds have been synthesized, 
often without full characterization, and only nitrides. 
Most of these are based on the Ca$^{2+}$ cation. 
All four of Ca$_3Pn_A$N were reported to be either semiconducting ($Pn_A$=Bi,Sb) 
or insulating ($Pn_A$=As,P).\cite{chern1992a}
Our calculated band gaps from Table \ref{table1} also indicate semiconducting states,
as did previous work on some of these members\cite{bilal2015b} and 
the Sr and Ba counterparts (Sr,Ba)$_3$(Sb,Bi)N.\cite{gaebler2004}
When Bi is substituted by the smaller pnictides Sb, As, and P, the band gap increases,
because as the electronegativity decreases, the energy level of the $p$ 
anion relative to the cation level decreases as well.

Structurally,
the $Pn_A$=Bi and Sb members of Ca$_3$$Pn_A$Bi are cubic, while (presumably) the size mismatch for the
smaller ions $Pn_A$=As and P anions
led to orthorhombic $Pnma$  distortion.\cite{chern1992b}
Niewa's review reported Sr$_3$(Sb,Bi)N as cubic,
as well as Mg$_3$(As,Sb)N\cite{chi2002} which are not included in our study.\cite{niewa2013}
In contrast, a hexagonal structure was reported for Ba$_3$(Sb,Bi)N, 
built on face-sharing rather than corner-linked octahedra.
Such a hexagonal structure tends to be favored by compounds containing alkaline-earth 
metal species with large radii.
Coulomb repulsion between N in face-sharing octahedra and the resulting 
distance d(N-N) has been suggested to play a role in stabilizing the 
structure.\cite{niewa2013,gaebler2004}
d(N-N) for Ba$_3$(Sb,Bi)N is 3.30\AA, which is sufficiently large to stabilize the hexagonal structure.

Some of these small gap compounds 
have been calculated to be promising
thermoelectric materials\cite{bilal2015,bilal2015b} with high thermodynamic figure of merit.
viz. Sr$_3$(Sb,Bi)N.\cite{bilal2015} 
Magnetic susceptibility and electrical resistivity data indicate
that the compounds are diamagnetic semiconductors.
The optical band gaps of Sr$_3$SbN and Sr$_3$BiN are reported to be
1.15 eV and 0.89 eV, respectively.\cite{gaebler2004}

First-principle calculations have been reported for some of the mentioned 
compounds, largely to assess
specific properties such as their elastic behavior and optical
response.\cite{haddadi2010,Rached2009,Jha2010,Ullah2016,Bidai2016,hichour2009b,bilal2014,haddadi2009c,hichour2010,moakafi2009,haddadi2009}
Our survey should be useful in interpretation of those results, by connecting their computed properties
to similarities and differences in their electronic structures.

\section{On the Possibility of Topological Character}
\label{topology}

\begin{figure}
\centering
\subfloat[$\mathrm{Ca_3BiN}$, a normal insulator with small gap.]
  {\includegraphics[width=\linewidth]{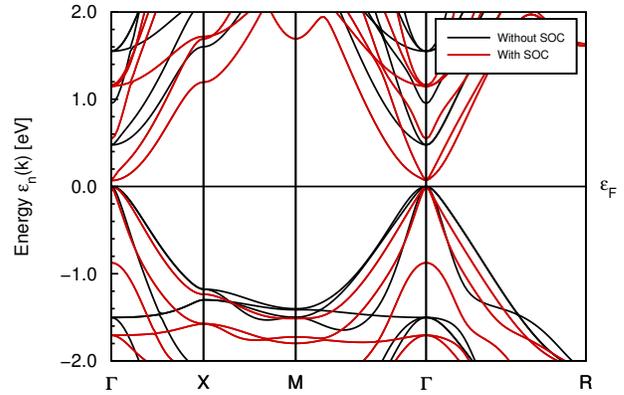}\label{ca3bin}}
\
\subfloat[$\mathrm{Ca_3BiP}$, a topological semimetal with CBM and VBM touching.]
 {\includegraphics[width=\linewidth]{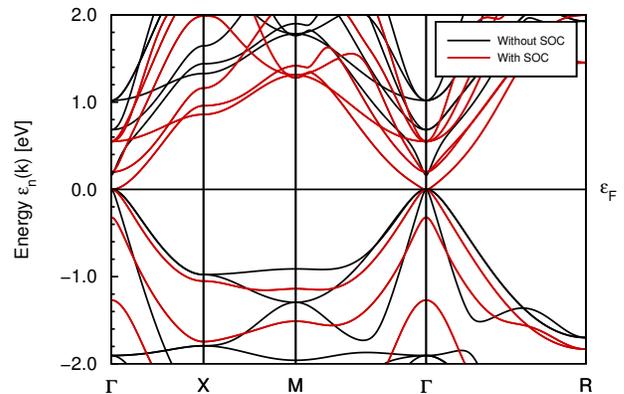}\label{ca3bip}}
\
\subfloat[$\mathrm{Sr_3BiP}$, a topological semimetal with delicate hole pockets,
  as indicated in the inset.]
 {\includegraphics[width=0.94\linewidth]{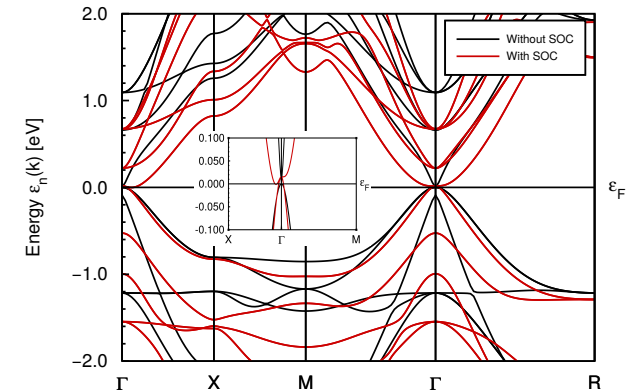}\label{sr3bip}}
\caption{The $Ae_3Pn_APn_B$ antiperovskite compounds can be classified into three
categories based on their electronic structure and topological invariants.
The electronic band structures of $\mathrm{Ca_3BiN}$, $\mathrm{Ca_3BiP}$ and
$\mathrm{Sr_3BiN}$, from top to bottom, are used as examples of each category.}
\label{categorize}
\end{figure}

The $Z_2$ invariant $\nu_0$ determines the topological nature of an insulator, and 
for inversion symmetric crystals the parity criteria proposed by Fu and Kane\cite{Fu2007}
determines the topological character.
Specifically, the sum of the parities $\delta_i$ at occupied states at the 
time-reversal invariant momenta (TRIM) determines the primary $Z_2$ invariant $\nu_0$:
\begin{equation}
 (-1)^{\nu_0}=\prod_{i=1}^8 \delta_i
\end{equation}

By studying Table \ref{table1} it can be found that the electronic structure and topological nature 
of these compounds can be classified into three categories:
(i) small gap with a topologically trivial phase, viz. $\mathrm{Ca_3BiN}$ in Fig. \ref{ca3bin},
(ii) the VBM and CBM touch at 
$\Gamma$ point with $Z_2$ indices of (1;000), as for $\mathrm{Ca_3BiP}$ in Fig. \ref{ca3bip}, and
(iii) there are (usually tiny)  electron and hole pockets along high symmetry 
lines with $Z_2$ indices of (1;000) as in the example of $\mathrm{Sr_3BiP}$ in Fig. \ref{sr3bip}.

For the first type, 
band inversion is not present for the cubic structure 
but the application of strain may lead to  inversion and a topological material.\cite{Sun2010}
For the second and third types (highlighted in bold in Table \ref{table1}), 
the band inversion is induced by SOC of the heavy elements.\cite{Jin2013} 
No gap is opened, however, leaving these systems as topological semimetals.
Strain can be used to induce a transition from
topological semimetal to topological insulator, provided that the band overlapping 
(especially for type (iii)) is not greater than the effect of SOC.

\begin{figure*}
\centering
\includegraphics[width=\textwidth]{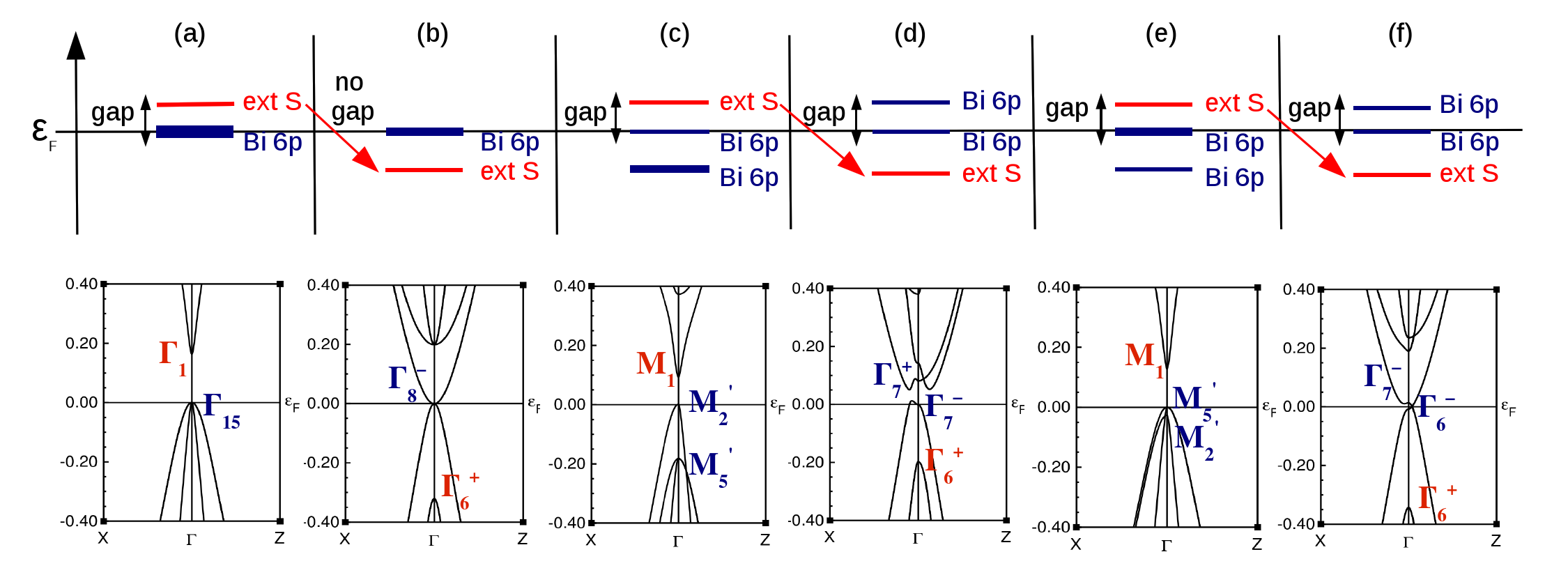}%
\caption{Upper panel: Schematic energy diagram of $\mathrm{Ca_3BiP}$ at $\Gamma$ point (a) without SOC, (b) with SOC, (c) with $5\%$ compressive strain and without SOC,
(d) with $5\%$ compressive strain and SOC, (e) with $1\%$ expansive strain and without SOC and (f) with $1\%$ expansive strain and SOC.
Ultra-fine, fine and thick lines represent 1-fold, 2-fold and 3-fold degeneracy (not including spin degeneracy) respectively.
Lower panel: Band structures along X -- $\Gamma$ -- Z with irreducible representations given (in Bouckaert-Smoluchowski-Wigner notation) at the $\Gamma$ point.}
\label{schematic}
\end{figure*}

A case study of $\mathrm{Ca_3BiP}$ will be used to demonstrate the role that interplay of SOC 
and strain play in band inversion of the second and third class above.
Fig. \ref{schematic} shows the schematic energy level diagram of $\mathrm{Ca_3BiP}$ 
at the $\Gamma$ point, along with the bands around $\Gamma$ for the same conditions.
Cubic $\mathrm{Ca_3BiP}$ has an energy gap of 0.2 eV.
At the $\Gamma$ point, the valence bands comprised of Bi $6p$ characters and P $3p$ 
characters have negative parities, while the $s$ character of the CBM has positive parity.
(There are substantial gaps at all other TRIMs, so their contributions to Z$_2$ are 
invariant to all changes and are the same for all of these antiperovskite compounds.)  
The strong SOC of Bi inverts the $s$ conduction band and the valence Bi $6p$ bands. 
No gap is opened, however, resulting in a topological semimetal phase.

With a tetragonal compression of 5$\%$ along (001), the degeneracy is lifted and a small 
gap of 34 meV is produced while maintaining the inverted band ordering. 
The strain does not violate inversion symmetry, but merely splits the degeneracy of the 
$\Gamma_8^-$ band at the Fermi level into two sets of Kramer doublets, 
with odd parity below the Fermi level and even parity above, so the parity eigenvalues are unchanged.
This mechanism is similar to that of $\mathrm{\alpha-Sn}$ and HgTe topological 
insulators.\cite{Fu2007,Ando2013}
Although HgTe does not have inversion symmetry, it can be argued that this compound in group II-VI can be treated as adding a inversion symmetry breaking perturbation into
group IV element, like Tin in the $\alpha$ or gray phase.

\begin{figure}
\centering
\includegraphics[width=0.4\textwidth]{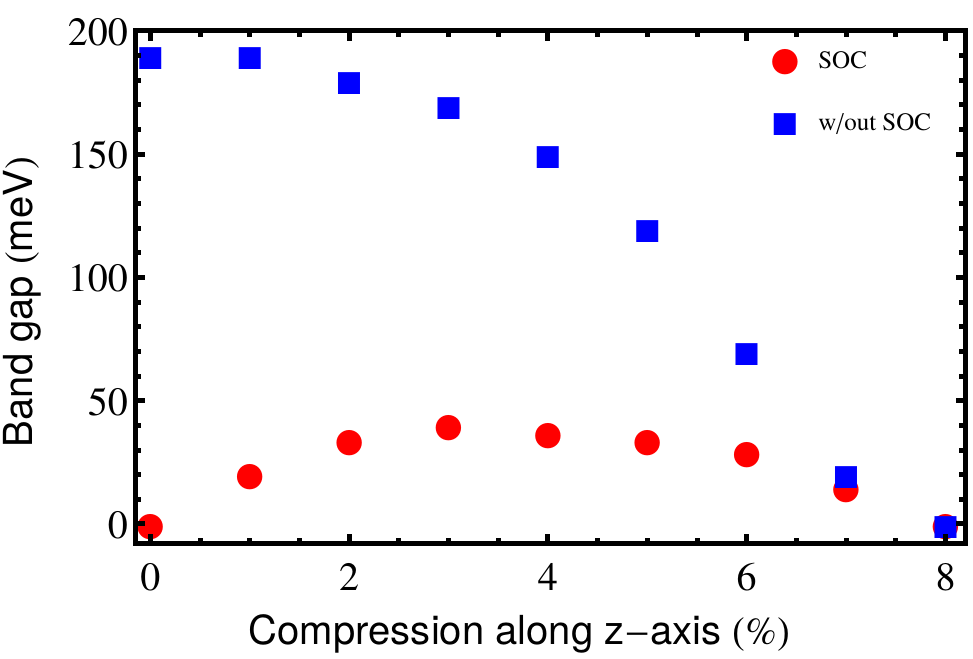}%
\caption{Band gap of Ca$_3$BiP versus compression, showing the effect of SOC.
Compression of along (001) without SOC closes the band gap. 
When SOC is included the band gap is closed, but then is re-opened by strain.}
\label{strain}
\end{figure}

The transition from the topological semimetal state to a topological insulator can also be 
realized by compressing the lattice parameter $c$ by a few percent, which can be accomplished
by growth of thin films on a substrate with the corresponding lattice parameter.
As shown in Fig.~\ref{strain}, a small energy band gap with SOC is produced within this range, 
with a maximum gap of 40 meV occurring at $3\%$ compression.
The band parity remain odd within this range, but goes back to even at 8$\%$ compression.
On the other hand, uniaxial expansion along the c-axis opens up a gap except along $k_z$, 
where a Dirac-like band crossing occurs, producing a topological Dirac 
semimetal (Fig.~\ref{schematic}f).
However, not all direction of strain application will open up a gap. 
For example, applying uniaxial strain along (111) direction does not open up a gap, because the Bi $p_x$, $p_y$ and $p_z$ characters still remain equivalent.

\section{Structural (In)stability}
\label{structure}

The oxide perovskite structure is notoriously subject to distortion from the ideal cubic 
structure, for two primary reasons. First, perovskite is not a close-packed structure. Especially
the BO$_2$ (001) layer (viz. the CuO$_2$ layer in high temperature superconducting cuprates)
is not: atoms are aligned in rows with an empty site of square symmetry. Second, there is tension
between the preferred lattice constants of the BO$_2$ layer and the AO layer. The mismatch is
quantified by Goldschmidt's ratio, using established ionic radii and quantifying the mismatch
as a percentage. It may be expected that these antiperovskites being studied may be subject
to similar issues. 
Unfortunately, these antiperovskites
comprise a new class of compounds for which appropriate ionic radii have not been established.  

\begin{figure}
 \subfloat[Phonon modes of $\mathrm{Ca_3BiN}$]{\includegraphics[width=0.44\textwidth]{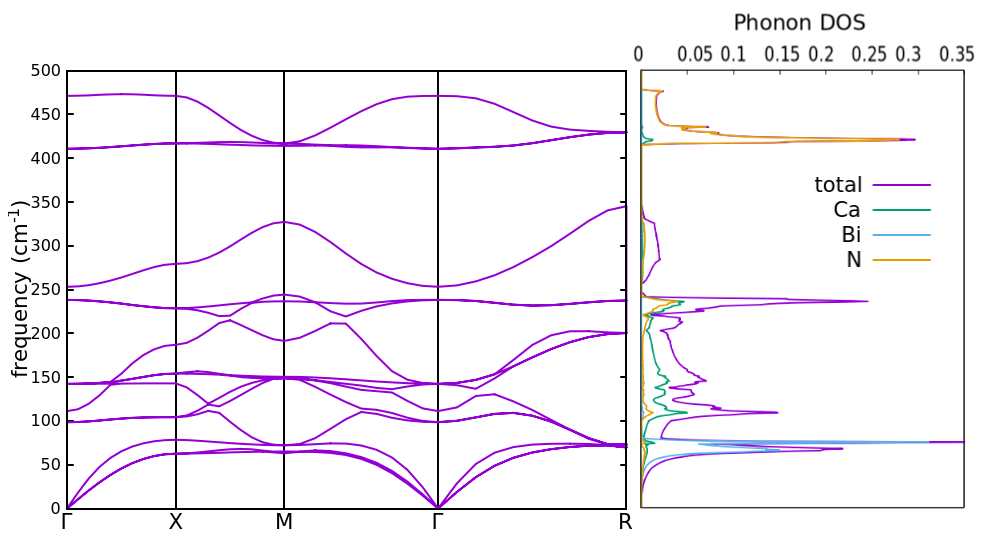}}\
 \subfloat[Phonon modes of $\mathrm{Ca_3BiP}$]{\includegraphics[width=0.44\textwidth]{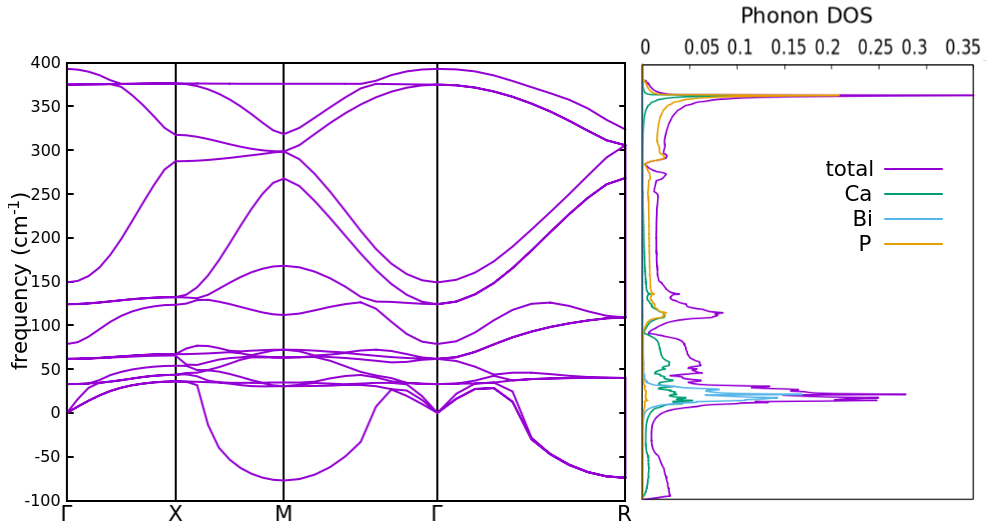}}     
\caption{Calculated phonon dispersion of (a) Ca$_3$BiN and (b) Ca$_3$BiP. 
Imaginary frequencies (plotted as negative) indicate structural instability, which is
severe for Ca$_3$BiP for both M and R point rotations.}    
\label{phonon} 
\end{figure}

Calculation of the phonon spectrum is commonly used to assess dynamical stability of a compound.
Antiperovskites with N atom at the B-site have been studied both experimentally and 
computationally
\cite{bilal2015,bilal2015b,chern1992b,niewa2013,gaebler2004,haddadi2010,Rached2009,Jha2010,Ullah2016,Bidai2016,hichour2009b,bilal2014,haddadi2009c,hichour2010,moakafi2009,haddadi2009}.
However, B-site cations other than N have not yet been incorporated into this structure.
We have calculated the phonon dispersion curves for 
 $\mathrm{Ca_3BiN}$ and for $\mathrm{Ca_3BiP}$
to assess their stability. The results are presented in Fig. \ref{phonon}.

\begin{figure}
\includegraphics[width=0.2\textwidth]{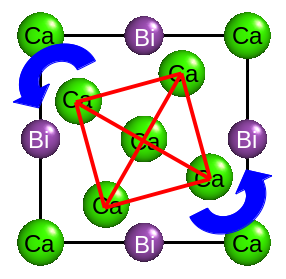}
\caption{Structure of distorted $\mathrm{Ca_3BiP}$ viewed along the z-axis, after the 
Ca octahedron (marked by red) has rotated around the z-axis.}
\label{bipca3-distort}
\end{figure}

For $\mathrm{Ca_3BiN}$, all frequencies are positive and the Bi atom dominates the acoustic
frequencies up to 75 cm$^{-1}$.
Ca atoms dominate frequencies in the $100 - 300$ cm$^{-1}$ range, 
while the light N atom oscillates in the highest frequency $400-475$ cm$^{-1}$ region.
The phonon spectrum of Ca$_3$BiP is quite different. The overriding feature is the
imaginary frequencies around both the M and R points of the zone.
The atomic displacements at these points give insight into the structural instability; the unstable
modes have Bi+Ca character.
At the M point, the Bi and P atoms are fixed while the Ca atoms rotate around the principal axis,
thus the crystal is unstable to such rotations.
Allowing the octahedron to rotate around z-axis as shown in Fig. \ref{bipca3-distort},
the energy is lowered by 0.2 eV.
The electronic rearrangement due to this distortion leaves  $\mathrm{Ca_3BiP}$ 
as a conventional insulator.

\begin{figure}
\includegraphics[width=0.45\textwidth]{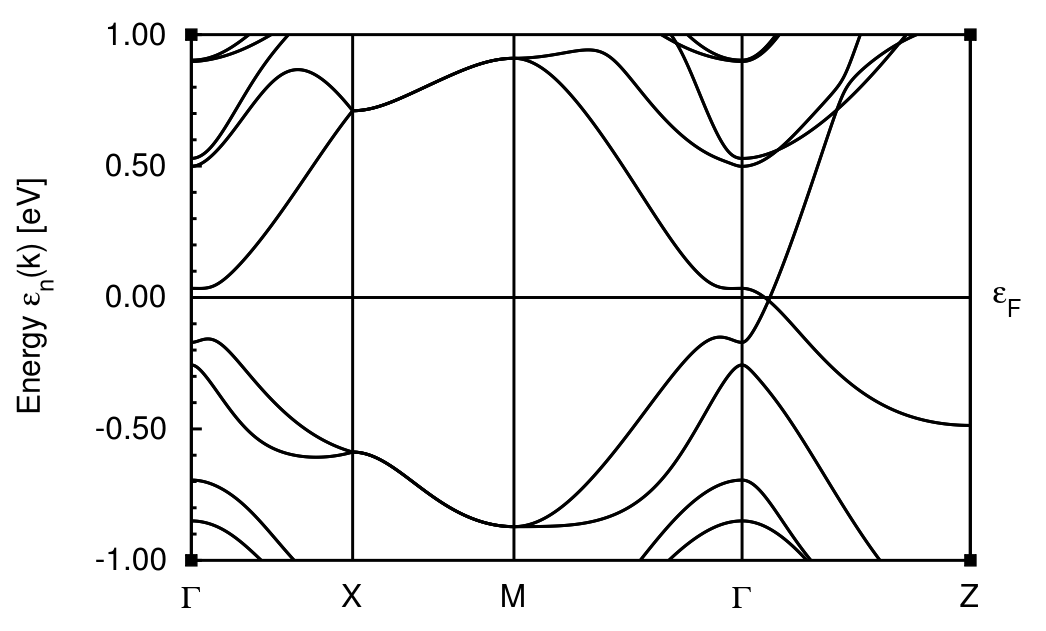}
\caption{Band structure of distorted $\mathrm{Ca_3BiSb}$ showing the Dirac 
point along $\Gamma-Z$ that pins the Fermi level.}
\label{bisbca3-distort}
\end{figure}

Distortion of antiperovskites compounds with heavier $Pn_B$ elements, 
viz. $Ae_3$BiSb with
$Ae$=Ca, Sr, and Ba, were studied as well. 
Distortions from cubic structure were favored by $0.7-0.8$ eV per formula unit.
The resulting band structures are those of a Dirac semimetal, hosting Dirac 
points along the $\Gamma-Z$ direction as shown for Ca$_3$BiSb in Fig.~\ref{bisbca3-distort}.
However, applying the mBJ correction widens that gap to around 0.4 eV and leaves them  
in conventional insulator states.

\section{Summary}
\label{summary}

The stability, electronic structure and topological aspect of the class of isovalent antiperovskite alkaline earth-pnictide compounds have been studied using DFT.
This class with pnictide other than N inside the octahedron is distorted with the octahedron rotated along a principal axis.
The electronic structures of these compounds with SOC can be classified into three categories.
First, one common class of the topological aspect is that the electronic structure is gapped with a topologically trivial phase.
Second, a zero gap semiconductor with VBM and CBM touching at $\Gamma$ point with a $Z_2$ invariant of 1;000.
Third, a semimetal consists of electron or hole pockets of maximum 0.1 eV energy and band degenerate at $\Gamma$ point with a $Z_2$ index of 1;000.
Strain is required to produce a topological insulator.
While the first type (e.g. $\mathrm{Ca_3BiN}$ \cite{Sun2010}) needs both SOC and proper strain to have band ordering inverted,
the second and third types (e.g. $\mathrm{Ca_3BiP}$ and $\mathrm{Sr_3BiP}$) only need SOC to induce the band inversion, giving a $Z_2$ invariant of 1;000, but in a topological semimetal state,
whereupon compressive strain may produce a transition to topological insulator.
On the other hand, expansive strain may give rise to Dirac semimetals.
With proper strain engineering, some may become a promising topological insulator and Dirac semimetal, as witness in $\mathrm{Ca_3BiP}$ and $\mathrm{Sr_3AsP}$ antiperovskite compounds.

\section{Acknowledgments}
We acknowledge useful comments from Theo Siegrist and Mas Subramanian about
pnictide-based antiperovskite and nitride perovskite synthesis. 
This work was supported by the NSF DMREF program through 
grant DMR-1534719. Computational data for the compounds in Table I was 
uploaded to the DOE-supported Materials Project.

\bibliographystyle{unsrt}

\end{document}